\documentclass[%
 reprint,
 amsmath,amssymb,
 aps,
]{revtex4-2}
\usepackage[utf8]{inputenc}
\usepackage{graphicx}
\usepackage{multirow}
\usepackage{dcolumn}
\usepackage{bm}
\usepackage{makecell}
\begin{document}

\preprint{APS/123-QED}

\title{Probing new physics at the LHC with $\mathrm{b}\tau\nu$ final states}

\author{Andrés Florez}%
 \email{ca.florez@uniandes.edu.co}
\affiliation{%
 Universidad de Los Andes, Cra. 1 \# 18a-12, Bogotá, Colombia
}%
\author{Tomas Atehortua Garces}
 \email{tomas.atehortua@udea.edu.co}
\author{José D. Ruiz Álvarez}%
 \email{josed.ruiz@udea.edu.co}
\affiliation{%
Instituto de Física, Universidad de Antioquia, A.A. 1226, Medellín, Colombia
}%





\date{\today}

\begin{abstract}
The $R_{D^{(\ast)}}$ anomaly is one of the most intriguing experimental results in particle physics today. Experiments such as BaBar, Belle and LHCb have measured a consistent tension with the standard model (SM). We study several extensions of the SM  that could potentially explain this tension, such as production of heavy $\mathrm{W}^{\prime}$ bosons, under the sequential SM scenario, leptoquarks with preferential couplings to third generation fermions, and interpretations through effective field theories. Such models, are not only able to explain the $R_{D^{(\ast)}}$ anomaly but also to produce distinctive signatures at the LHC. We present different feasibility studies to probe each of these scenarios at the LHC, considering final states with one $\mathrm{b}$-quark candidate, one hadronically decaying tau lepton ($\tau_{h}$) and missing transverse momentum ($p^{miss}_{\mathrm{T}}$). The selection criteria has been optimized for each model to achieve best signal signficance. The studies are performed considering different LHC running conditions, at $\sqrt{s} = 13 \, \mathrm{TeV}$ and 13.6 $\mathrm{TeV}$, and different luminosities (150~$\mathrm{fb}^{-1}$ and 3000~$\mathrm{fb}^{-1}$). 
\end{abstract}

\maketitle


\section{\label{sec:level1}Introduction}

The Standard Model (SM) of particle physics synthesizes our understanding of fundamental particles and their interactions. Although it has been a successful theory to explain a broad set of experimental results, it does not provide answers to several questions raised  at the theoretical and experimental level. With the aim to elucidate some of these conundrums, the LHC has a broad physics program to search for new particles that could explain some of the insufficiencies of the SM. A plethora of new models have been proposed, suggesting the existence of new particles such as heavy gauge bosons, neutral ($\mathrm{Z}'$) or electrically charged ($\mathrm{W}'$)~\cite{Mohapatra:1985xm, Lykken:1996kz}, Leptoquarks ($\mathrm{LQ}$)~\cite{Schmaltz:2018nls}, heavy neutrinos~\cite{gluza2015heavy, astier2001search, aad2015search}, new partners of the Higgs boson~\cite{falkowski2015second, celis2013lhc, farzinnia2014higgs}, and others. Many of these new ideas have been experimentally probed by the ATLAS~\cite{ATLAS:2008xda} and CMS~\cite{CMS:2008xjf} experiments at the LHC, without positive results until now. 

Experimental searches have used a variety of models as benchmark scenarios to set upper limits on the production cross section of the hypothetical particles, at 95\% C.L, for a broad range of masses and couplings. For example, several searches for $\mathrm{Z}'$~\cite{CMS:2021ctt, ATLAS:2019erb, cms2012search} and $\mathrm{W}'$ \cite{cms2015search, cms2018search, atlas2011search} gauge bosons have set their exclusion ranges considering scenarios where these particles have the same couplings as those of the SM $\mathrm{Z}$ and $\mathrm{W}$ bosons. This particular benchmark model is known as the sequential SM (SSM). Existing limits to third generation fermions are less constrained in SSM interpretations and remain important to test lepton flavor universality (LFU). In the case of $\mathrm{LQ}$s, the searches have been performed considering democratic coupling scenarios to all SM fermions or enhanced couplings to second or third generation fermions~\cite{CMS:2020wzx}. 

Although we currently do not have conclusive experimental evidence of physics beyond the SM, there is a set of tensions on the decay ratios of $\mathrm{B}$ mesons, $R^{\tau/\ell'}_{D^{(*)}} (\ell'=\mathrm{e},\mu$), $R^{\tau/\ell'}_{D}$, $R^{\mu/\mathrm{e}}_{K^{(*)}}$, and $R^{\mu/\mathrm{e}}_{K}$, measured by the BaBar~\cite{BaBar:2015wkg, Lees:2013uzd, Lees:2012xj, BaBar:2006tnv}, Belle~\cite{Abdesselam:2019dgh, Hirose:2017dxl, Belle:2016fev, Sato:2016svk, Hirose:2016wfn, Belle:2016xuo,  Huschle:2015rga, Belle:2009zue}, and LHCb collaborations~\cite{LHCb:2019hip, Aaij:2017uff, LHCb:2016ykl, LHCb:2015wdu, Aaij:2015yra, LHCb:2014cxe}. To explain these experimental observations, some theoretical models and phenomenological studies propose scenarios where new particles have non-universal/preferential couplings to third generation fermions \cite{Marzocca:2020ueu, Endo:2021lhi, Gomez:2019xfw, Iguro:2020keo, Iguro:2018fni}. 

We perform a feasibility study to assess the long-term discovery potential for resonant production of $\mathrm{W}'$ bosons and $\mathrm{LQ}$s at the LHC, considering preferential couplings to third generation fermions, motivated by the $\mathrm{B}$ meson anomalies. We also consider a scenario where these hypothetical mediators are so heavy they can only be produced in off-shell processes. This approach scenario is studied with an effective field theory (EFT) formulation. Following \cite{abdullah2018probing}, we explore the production of $\mathrm{W}'$ bosons that couple to bottom ($\mathrm{b}$) and charm ($\mathrm{c}$) quarks on the color sector but have democratic couplings to SM leptons (SSM): $\mathrm{pp} \rightarrow \mathrm{W}'+\mathrm{b/c}$. We consider final states where the $\mathrm{W}'$ decays to a tau lepton, which subsequently decays via hadronic modes ($\tau_{h}$) and associated neutrinos ($\mathrm{W}' \rightarrow \tau \nu_{\tau} \rightarrow \tau_{h} \nu_{\tau}  \nu_{\tau} $). In addition, we also explore single production of $\mathrm{LQ}$s, decaying to a $\tau$ lepton ($\tau_{h}$) and a $\mathrm{b}$-quark ($\mathrm{LQ} \rightarrow \tau \mathrm{b} \rightarrow \tau_{h} \nu_{\tau} \mathrm{b}$). Therefore, our final state is composed of a $\tau_{h}$, a $\mathrm{b}$-quark jet and missing transverse momentum ($p_{\mathrm{T}}^{miss}$) from associated neutrino(s). The $p_{\mathrm{T}}^{miss}$ variable is defined as the magnitude of the negative vectorial sum of all reconstructed objects by the detector. This quantity is an indirect measurement of the momentum of particles that escape detection, such as neutrinos. The EFT formulation considers an effective coupling between the $\mathrm{b}$ and $\mathrm{c}$ quarks, the $\tau$ and $\nu_{\tau}$. Although we target the same final state ($\mathrm{b}$-quark, $\tau_{h}$, $p_{\mathrm{T}}^{miss}$), the topology and observables vary depending on the model. For example, the $\mathrm{b}$-quark in the $\mathrm{LQ}$ scenario is significantly more energetic with respect to the same object in the $\mathrm{W}'+\mathrm{b/c}$ interpretation. This affects the expected experimental sensitivity in each case.

The studies performed within different models are carried out considering proton-proton ($\mathrm{pp}$) collisions at $\sqrt{s} = 13$ $\mathrm{TeV}$ and 13.6 $\mathrm{TeV}$. Different LHC luminosity scenarios are assumed. The estimation of the experimental sensitivity for each benchmark scenario is performed for different sets of masses and couplings. We determine the relevant variables and optimal thresholds to maximize experimental sensitivity.

\section{Theoretical models}

In the SSM, an additional $\text{SU}(2)'$ symmetry group is proposed, which leads to three additional gauge bosons, one neutral, $\mathrm{Z}'$, and two electrically charged $\mathrm{W}'^\pm$~\cite{barger1980sequential}. Depending on the realization of the model, different couplings to leptons are considered. The Lagrangian for the charged current is defined in Equation~\ref{eq:lagssm}, where $\mathrm{CC}$ stands for charged current and the $g_2'$ factor is the coupling constant for the additional $\text{SU}(2)'$ group. The $\kappa$ parameters are complex values of $3\times3$ matrices in the flavour space.

\begin{gather}
    \mathcal{L}_{\mathrm{CC}}^{W^{\prime}}=\frac{g_{2}^{\prime}}{\sqrt{2}}\bar{u}_{i} \gamma^{\mu}\left(\left[\kappa_{q, L}^{W^{\prime}}\right]_{i j} P_{L}+\left[\kappa_{q, R}^{W^{\prime}}\right]_{i j} P_{R}\right) d_{j}+ \nonumber \\
    \frac{g_{2}^{\prime}}{\sqrt{2}}\bar{\nu}_{i} \gamma^{\mu}\left(\left[\kappa_{\ell, L}^{W^{\prime}}\right]_{i j} P_{L}+\left[\kappa_{\ell, R}^{W^{\prime}}\right]_{i j} P_{R}\right) \ell_{j} W_{\mu}^{\prime}+\text { h.c. }
    \label{eq:lagssm}
\end{gather}

The structure of the Lagrangian is similar to that of the SM, and in fact the SM can be recovered by performing the replacements outlined in Equation~\ref{eq:ssmrepl} below and summing over quark and lepton flavors, as described in~\cite{araz2021crossfertilising}. In this model, the $\mathrm{W}'$ and the $\mathrm{Z}'$ masses are free parameters.

\begin{gather}
{\left[\kappa_{q, L}^{W^{\prime}}\right]_{i j}=V_{i j}^{\text{CKM}}, \quad\left[\kappa_{\ell, L}^{W'}\right]_{i j}=V_{i j}^{\text{PMNS}}, \quad \kappa_{q(\ell), R}^{W^{\prime}}=0} \nonumber\\
\left[\kappa_{f, L}^{Z^{\prime}}\right]_{i j}=\left(T_{L}^{3, f}-Q_{f} \sin ^{2} \theta_{W}\right) \delta_{i j}, \nonumber\\ 
\quad\left[\kappa_{f, R}^{Z'}\right]_{i j}=\left(-Q_{f} \sin ^{2} \theta_{W}\right) \delta_{i j}
\label{eq:ssmrepl}
\end{gather}

In addition to the $\mathrm{W}^{\prime}$/$\mathrm{Z}^{\prime}$ model described above, we consider models with $\mathrm{LQs}$ as massive intermediate particles, with fractional electric charge, that couple simultaneously to leptons and quarks~\cite{Schmaltz:2018nls}. These hypothetical particles could be of scalar or vectorial nature, depending on the model, and their mass is a free parameter. As an example,  Equation~\ref{eq:vlqlgra} shows the Lagrangian for a vectorial $\mathrm{LQ}$ defined as $U_{1}$, with quantum numbers ($3_{C}, 1_{I}, 2/3_{Y}$), that couples to the second and third families of quarks and leptons. 

\begin{align}
\mathcal{L}_{U} &\supset U_{1,\mu}(\lambda_{33}\bar{q}_{3}\gamma^{\mu}P_{L}l_{3}+\lambda_{32}\bar{q}_{3}\gamma^{\mu}P_{L}l_{2}  \nonumber \\
 &+\lambda_{23}\bar{q}_{2}\gamma^{\mu}P_{L}l_{3}+\lambda_{22}\bar{q}_{2}\gamma^{\mu}P_{L}l_{2}) + \text { h.c. }
\label{eq:vlqlgra}
\end{align}

For our final model, we consider an EFT formulation. 
An EFT is an approximation that allows us to include degrees of freedom to describe phenomena occurring at a certain energy scale, without considering the substructure of the interactions at higher energies or other underlying physics phenomena~\cite{grozin2020effective}. An EFT only includes  relevant degrees of freedom, i.e., those states with $m \ll \Lambda$ while heavier excitations with $M \gg \Lambda$ are integrated out from the action~\cite{Manohar:1996cq, Manohar:2018aog, Brivio:2017vri, Buchmuller:1985jz, Falkowski:2017pss}. This means that the heavier degrees of freedom are placed in the ``low energy Lagrangian''. The representative Lagrangian for the EFT model considered in these studies is shown in Equation \ref{eq:etflagr}.

\begin{align}
\mathcal{L}_{\mathrm{eff}} \supset &-\frac{2 V_{i b}}{v^{2}}\left[\left(1+\epsilon_{L}^{i b}\right)\left(\bar{\tau} \gamma_{\mu} P_{L} \nu_{\tau}\right)\left(\bar{u}_{i} \gamma^{\mu} P_{L} b\right)\right.\nonumber\\
&+\epsilon_{R}^{i b}\left(\bar{\tau} \gamma_{\mu} P_{L} \nu_{\tau}\right)\left(\bar{u}_{i} \gamma^{\mu} P_{R} b\right)\nonumber\\
&+\epsilon_{T}^{i b}\left(\bar{\tau} \sigma_{\mu \nu} P_{L} \nu_{\tau}\right)\left(\bar{u}_{i} \sigma^{\mu \nu} P_{L} b\right)\nonumber\\
&\left.+\epsilon_{S_{L}}^{i b}\left(\bar{\tau} P_{L} \nu_{\tau}\right)\left(\bar{u}_{i} P_{L} b\right)\right]+\text { h.c. }
\label{eq:etflagr}
\end{align}
 
\begin{figure}[h]
\includegraphics[width=7cm]{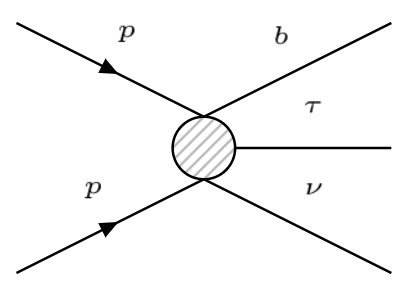}
\caption{Feynman diagram of $\textrm{pp}\to \textrm{b}\tau\nu$ production for the EFT model. The circle in the middle represents the new physics that could explain the $R_{D^{\ast}}$ anomaly. We don't specify the nature of the final state particles, as particles or their antiparticles, as we are interested in all the possible final states.}
\label{fig:EFTfeyfiag}
\end{figure}

 In Equation~\ref{eq:etflagr}, $V_{ij}$ are the Cabibbo-Kobayashi-Maskawa (CKM) matrix elements, $P_{R,L}$ are the chiral projectors, $\sigma^{\mu\nu} = (i/2)[\gamma^{\mu},\gamma^{\nu}]$, $v\approx 246$~$\mathrm{GeV}$ is the electroweak symmetry breaking scale, and $\epsilon_\Gamma$ are the Wilson Coefficients (WC)~\cite{greljo2019mono}. Figure \ref{fig:EFTfeyfiag} shows a representative Feynman diagram of $\textrm{pp}\to \textrm{b}\tau\nu$ production under the EFT interpretation.\\ 
 
\section{Simulated samples}

Signal and background samples are generated at parton level using MadGraph5\_aMC (v2.6.3.2)~\cite{Alwall:2014hca}, considering $\mathrm{pp}$ collisions with a center-of-mass energy of $\sqrt{s} = 13$~$\mathrm{TeV}$ and 13.6 $\mathrm{TeV}$. For the parton distribution function (PDF), the NNPDF3.0 NLO~\cite{NNPDF:2014otw}  set is used in the event generation of all simulated samples. At this stage, jets are required to have a minimum transverse momentum ($p_{T}$) of 30~$\mathrm{GeV}$ and $|\eta| < 5.0$. Generated partonic events are then passed through the PYTHIA 8 (v8.2.05)~\cite{Sjostrand:2014zea} package to include hadronic processes. The  DELPHES (v3.4.1)~\cite{deFavereau:2013fsa} software is used to emulate detector response, using the CMS detector configuration, for particle reconstruction and identification efficiencies. 

The MLM algorithm~\cite{Alwall:2007fs} is used for jet matching and jet merging. This process is used for the simulation of backgrounds which include additional jets. We follow the common recommendations for this procedure. 

The dominant sources of background events come from the production of a SM $\mathrm{Z}$ ($\mathrm{W}$) gauge boson with associated jets, referred to as $\mathrm{Z}+$jets ($\mathrm{W}+$jets), and the production of pairs of top quarks with additional jets ($\mathrm{t}\bar{\mathrm{t}}$). Events from single production of top quarks plus jets, gauge boson pairs ($\mathrm{WW}$, $\mathrm{ZZ}$, $\mathrm{WZ}$), and from Quantum Chromodynamic interactions among light quarks (QCD multijets) are also considered but deemed negligible after applying the main event selection criteria to search for signal. Based on the final state objects used in the study, $2\times 10^{6}$ events are produced for each background, considering leptonic decays for the $\mathrm{Z}+$jets ($\mathrm{W}+$jets) background and semileptonic final states for $\mathrm{t} \bar{\mathrm{t}}$.

The production cross sections for all simulated processes are estimated using MadGraph5 at leading order precision. Table~\ref{tab:crosssec} shows the cross sections we use for the different signal models and Table~\ref{tab:xsbkg} for the background samples.     

\begin{table*}[]
\centering
\begin{tabular}{c c c c}
\hline
Model                     & Parameters                         & Cross Section $13\,\text{TeV}$ [$\mathrm{pb}$]  &  Cross Section $13.6 \,\text{TeV}$ [$\mathrm{pb}$]\\ \hline\hline
\multirow{6}{*}{SSM}      & $m_{W'} = 6.0\times10^2$ $\mathrm{GeV}$      & $5.25$ & $5.88$\\ 
                          & $m_{\mathrm{W}'} = 1.0\times10^3$ $\mathrm{GeV}$      & $0.45$ & $0.52$\\
                          & $m_{\mathrm{W}'} = 1.6\times 10^3$ $\mathrm{GeV}$   & $3.04\times10^{-2} $& $3.45\times10^{-2}$\\ 
                          & $m_{\mathrm{W}'} = 2.0 \times 10^3$ $\mathrm{GeV}$   & $1.06\times 10^{-2}$ & $1.26\times 10^{-2}$\\
                          & $m_{\mathrm{W}'} = 2.5 \times 10^3$ $\mathrm{GeV}$   & $3.11\times 10^{-3}$ & $3.58\times 10^{-3}$\\
                          & $m_{\mathrm{W}'} = 3.0 \times 10^3$ $\mathrm{GeV}$   & $1.18\times 10^{-3}$ & $1.36\times 10^{-3}$\\ \hline
\multirow{3}{*}{EFT}      & $\epsilon_l^{cb}= 1.0$             & $0.13$ & $0.14$\\ 
                          & $\epsilon_{sL}^{cb}= 1.0$          & $8.07\times10^{-2}$ & $9.08\times10^{-2}$\\  
                          & $\epsilon_t^{cb}= 1.0$             & $0.71$ & $0.79$ \\ \hline
\multirow{6}{*}{LQ\_U(1)} & $m_{\mathrm{LQ}} = 5.0\times 10^2$ $\mathrm{GeV}$   & $0.74$ & $0.93$\\ 
                          & $m_{\mathrm{LQ}} = 1.0\times 10^3$ $\mathrm{GeV}$   & $2.44\times10^{-2}$ &$2.94\times10^{-2}$ \\  
                          & $m_{\mathrm{LQ}} = 1.25\times 10^{3}$ $\mathrm{GeV}$ & $6.80\times 10^{-2}$ & $8.56\times10^{-3}$ \\
                          & $m_{\mathrm{LQ}} = 1.5\times 10^3$ $\mathrm{GeV}$   & $2.20\times 10^{-3}$ & $2.77\times10^{-3}$ \\ 
                          & $m_{\mathrm{LQ}} = 2.0\times 10^3$ $\mathrm{GeV}$   & $3\times 10^{-4}$ & $3.94\times10^{-4}$ \\  
                          & $m_{\mathrm{LQ}} = 3.0\times 10^{3}$ $\mathrm{GeV}$ & $9.82\times 10^{-6}$ & $1.24\times10^{-5}$ \\ \hline
\end{tabular}
\caption{Signal parametrization and cross sections.}
\label{tab:crosssec}
\end{table*}

\begin{table*}[]
\centering
\begin{tabular}{c c c c}
\hline
Process    & Monte Carlo restrictions                            & Cross Section $13\,\text{TeV}$[pb] & Cross Section $13.6\,\text{TeV}$[pb] \\ \hline\hline
$\mathrm{t}\Bar{\mathrm{t}}$ & --                                    & 504.0 & 558.9 \\ 
$\mathrm{W} +jets$  & \texttt{ptl} $= 190$ $\mathrm{GeV}$, \texttt{misset} $= 160$ $\mathrm{GeV}$ & $0.64$ & $0.71$\\ 
$DY + jets$ & \texttt{ptl} $= 190$ $\mathrm{GeV}$ & $0.25$ & $0.27$\\ \hline
\end{tabular}
\caption{Production cross section for dominant backgrounds.}
\label{tab:xsbkg}
\end{table*}

Signal samples are produced as $\textrm{pp}\to \textrm{b}\tau\nu$  without additional jets, using a Feynrules~\cite{Alloul:2013bka} implementation interfaced with MadGraph in the UFO format~\cite{Degrande:2011ua}. We use the same models utilized in~\cite{greljo2019mono}. For the $\mathrm{W}'$ model, we only consider couplings to $\mathrm{b}$ and $\mathrm{c}$ quarks, $g_{q}$, as well to $\tau$ and $\nu_{\tau}$, $g_{l}$. The simulated samples are produced considering $g_{q} = g_{l} = 1.0$ and masses of 0.6, 1.0, 1.6, 2.0, 2.5, 3.0, and 3.5~$\mathrm{TeV}$, with fifty thousand events per sample. For the $\mathrm{LQ}$ case, we produce samples with $g_{bl} = g_{cl} = 1.0$ (couplings of the LQ and the b or quark and leptons) and masses of 0.5, 1.0, 1.25, 1.5, 2.0, and 3.0~$\mathrm{TeV}$, with one hundred thousand events per sample. For the $\mathrm{W}'$ and $\mathrm{LQ}$ samples, the corresponding decay widths are left to be automatically calculated by MadGraph. For the EFT scenario we consider separately the case with a scalar, vectorial, and tensor coupling. For each case we produce a Monte Carlo sample, considering a coupling of 1.0 and one hundred thousand events per sample. 

\section{Event selection criteria}

The event selection criteria is divided in two sets. The first set, defined as baseline selections, contain the  requirements used similarly across the tree different models under study. These selections allow us to define the phase space and objects to search for each type of signal. Events with one $\tau_{h}$ candidate, zero electrons or muons, and exactly one $\mathrm{b}$-quark jet are selected, following the studies performed in reference~\cite{abdullah2018probing}. Events with two or more $\tau_{h}$ candidates with $p_{\mathrm{T}}$ above 50 $\mathrm{GeV}$ and $|\eta_\tau| < 2.3$ are rejected. The baseline selections are summarized in Table~\ref{tab:baseline}. The second set of selections, presented in Table~\ref{tab:topocuts}, are associated with  the topological characteristics of each model. These selections and the corresponding thresholds have been chosen using a $\frac{N_{s}}{\sqrt{N_{s}+ N_{b}}}$ figure of merit, in order to obtain the best signal significance, where $N_{s}$ ($N_{b}$) represents the expected number of signal (background) events, normalized to cross section and luminosity.

\begin{table}[]
    \centering
        \begin{tabular}{lr}\hline
        Criterion                                   & Selection                               \\\hline\hline
        $N(\tau_{h})$                               &
        $> 0$                                  \\
        $|\eta_\tau|$                               & $\leq 2.3$                             \\
        Veto $2^\text{nd}-\tau_{h}$                          & $p_\mathrm{T}>50$ $\mathrm{GeV}$ \& $|\eta| < 2.3$ \\
        $N_{e/\mu}$ with $p_T(e/\mu) > 15$ $\mathrm{GeV}$ & $ = 0$                                  \\
        $N_{\mathrm{b}-\text{jets}}$                           & $=$ 1                          \\
        $p_\mathrm{T}(\mathrm{b})$                                    & $>20$ $\mathrm{GeV}$                       \\
        $|\eta_{\mathrm{b}-\text{jets}}|$                      & $<2.5$                                  \\\hline
        \end{tabular}
        \caption{Baseline selection criteria used for the  different models under study.}
        \label{tab:baseline}
\end{table}

\begin{table}[h]
\begin{tabular}{c c c c}
\hline
Parameter            & SSM               & EFT       & U1 LQ         \\ \hline\hline
$p_{\mathrm{T}}(\tau) >$        & $250$ $\mathrm{GeV}$ & $200$ $\mathrm{GeV}$ & $300$ $\mathrm{GeV}$ \\ 
$|\Delta \phi (\tau,p_{\mathrm{T}}^\text{miss})|>$ & $1.5$             & $2.0$              & $1.0$         \\ 
$p_{\mathrm{T}}^\text{miss}>$                & $200$ $\mathrm{GeV}$ & $300$ $\mathrm{GeV}$  & $400$ $\mathrm{GeV}$ \\ 
\hline
\end{tabular}
\caption{Topological selections for the three different models, $\mathrm{W}'$, $\mathrm{LQ}$ and EFT, considered in the analysis. Thresholds have been selected based on best signal significance.}
\label{tab:topocuts}
\end{table}

\section{Results}

\begin{figure}[]
\includegraphics[width=8cm]{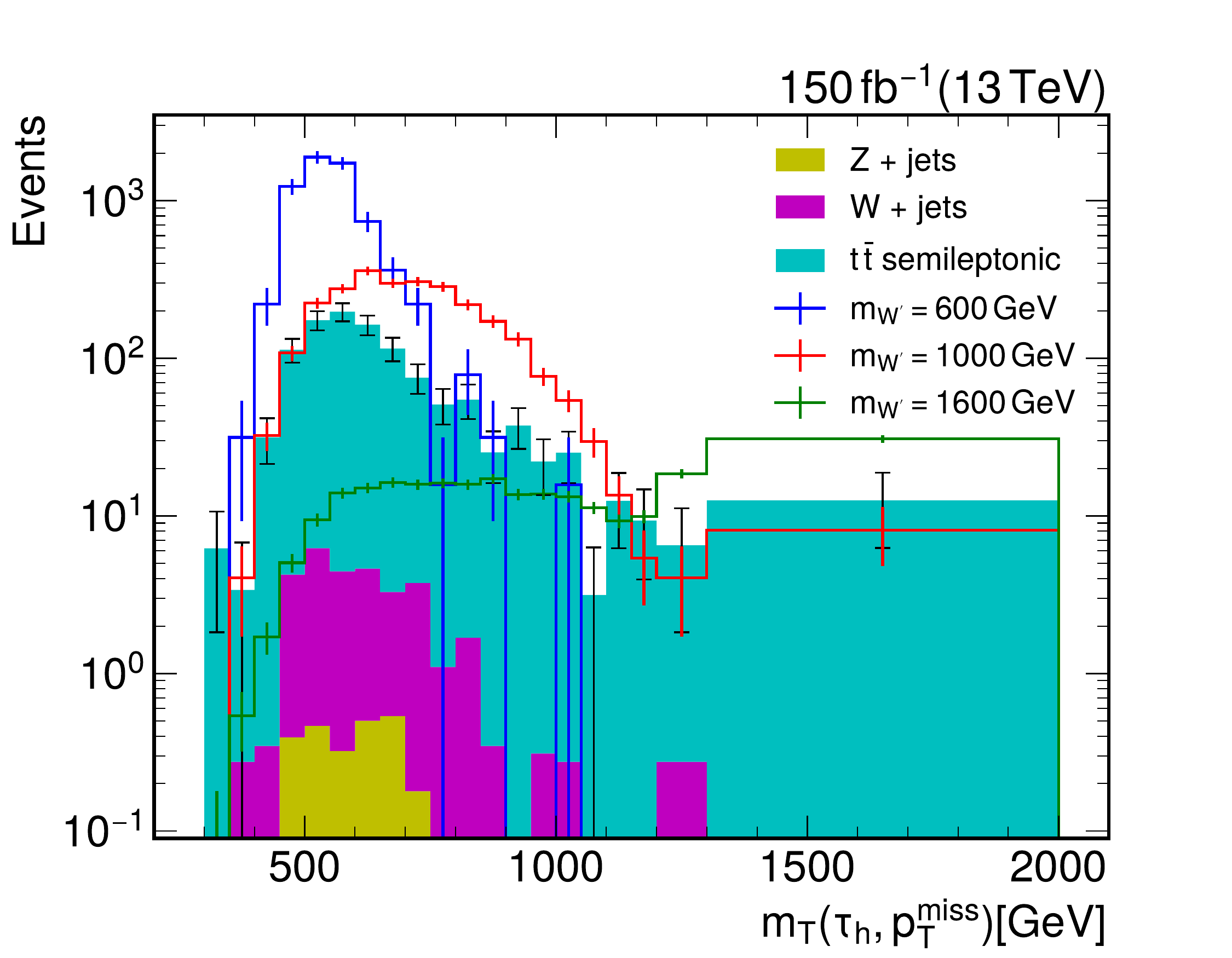}
\caption{Distribution of the $m_{\mathrm{T}} (\tau_{h}, p_{\mathrm{T}}^{miss})$ variable, for the background prediction and three different signal points. The backgrounds are represented by solid colors and are stacked on top of each other, while the signal samples are represented by solid lines and overlaid on top of the background. The expected events are estimated for an integrated luminosity of 150 $\mathrm{fb}^{-1}$ and $\sqrt{s} = 13$ $\mathrm{TeV}$.}
\label{fig:wprimemt}
\end{figure}

After selecting events passing the selection criteria outlined in Tables~\ref{tab:baseline} and~\ref{tab:topocuts}, three different observables, one per model, are chosen to assess the presence of 
signal events among the background expectation. These observables are found to give the best separation between signal and background events, maximizing the expected experimental sensitivity. For the $\mathrm{W}'$ scenario, the transverse mass between the $\tau_{h}$ candidate and expected missing transverse momentum, defined as  $m_{\mathrm{T}} (\tau_{h}, p_{\mathrm{T}}^{miss}) = \sqrt{2 p_{T}(\tau_{h}) p_{\mathrm{T}}^{miss} \times \cos (\Delta \phi (\tau_{h}, p_{\mathrm{T}}^{miss})) }$, is chosen. The $m_{\mathrm{T}} (\tau_{h}, p_{\mathrm{T}}^{miss})$ distribution is shown in Figure~\ref{fig:wprimemt}. The figure shows the background expectation in solid format, stacked, and three different signal samples overlaid on top of the background. Events are normalized based on the cumulative efficiency after all the selections, the production cross sections, and integrated luminosity of 150 $\mathrm{fb}^{-1}$. The choice of luminosity is based on the performance of the LHC during 2016-2018 for $\mathrm{pp}$ collisions at $\sqrt{s} = 13$ $\mathrm{TeV}$.    

\begin{figure}[]
\includegraphics[width=9cm]{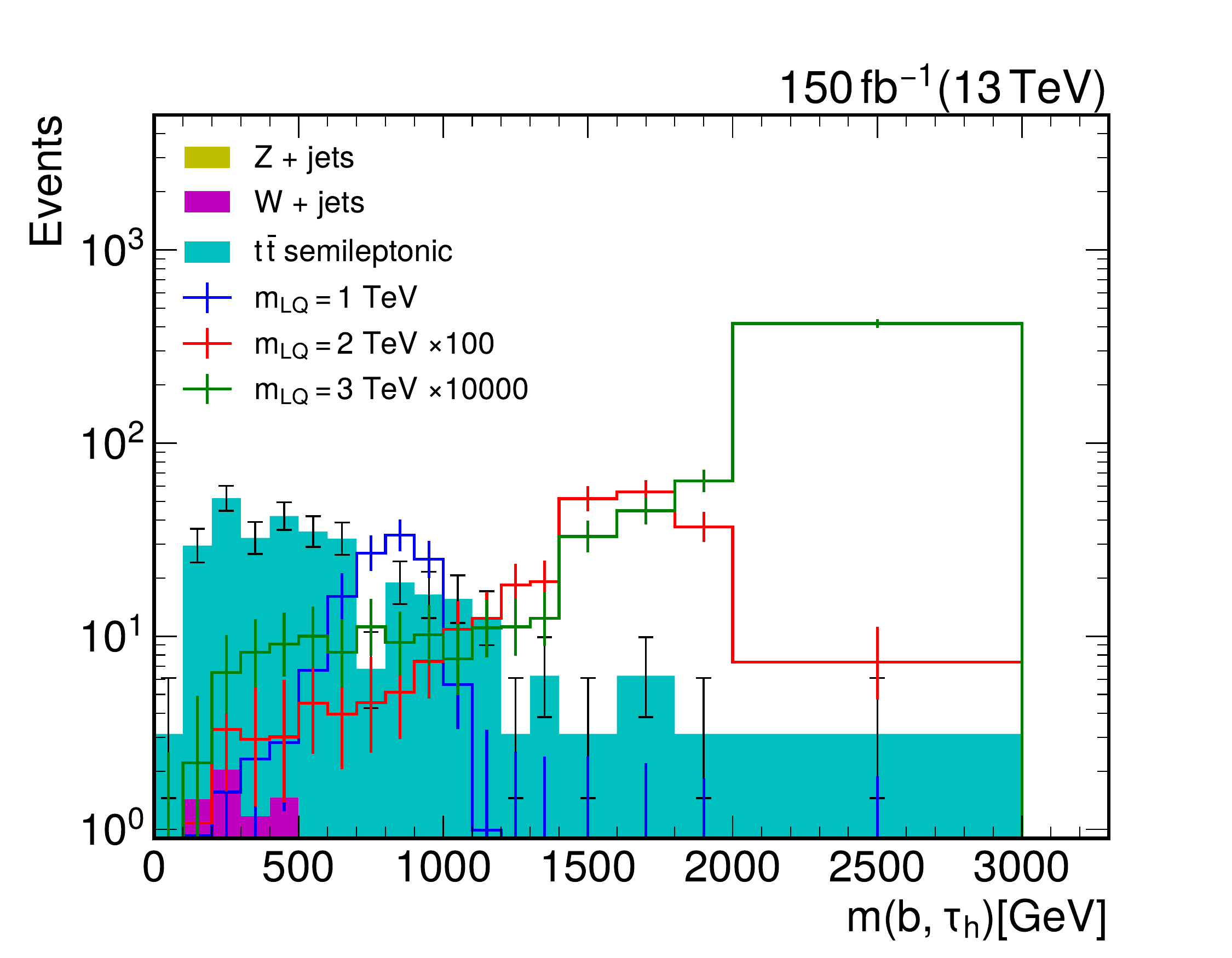}
\caption{Distribution of the reconstructed mass between the $\tau_{h}$ and $\mathrm{b}$-quark jet candidates. The backgrounds are represented by solid colors and are stacked while the signal samples are represented by solid lines and overlaid on top of the background. The expected event yields are estimated for an integrated luminosity of 150 $\mathrm{fb}^{-1}$ and $\sqrt{s} = 13$~$\mathrm{TeV}$.}
\label{fig:lqmass}
\end{figure}

For the $\mathrm{LQ}$ model, the reconstructed mass between the $\tau_{h}$ and $\mathrm{b}$ quark candidates is used as the main observable to search for signal. Figure~\ref{fig:lqmass} shows the corresponding distributions, following the same conventions as those used for $m_{\mathrm{T}} (\tau_{h}, p_{\mathrm{T}}^{miss})$. 

\begin{figure}[]
\includegraphics[width=8cm]{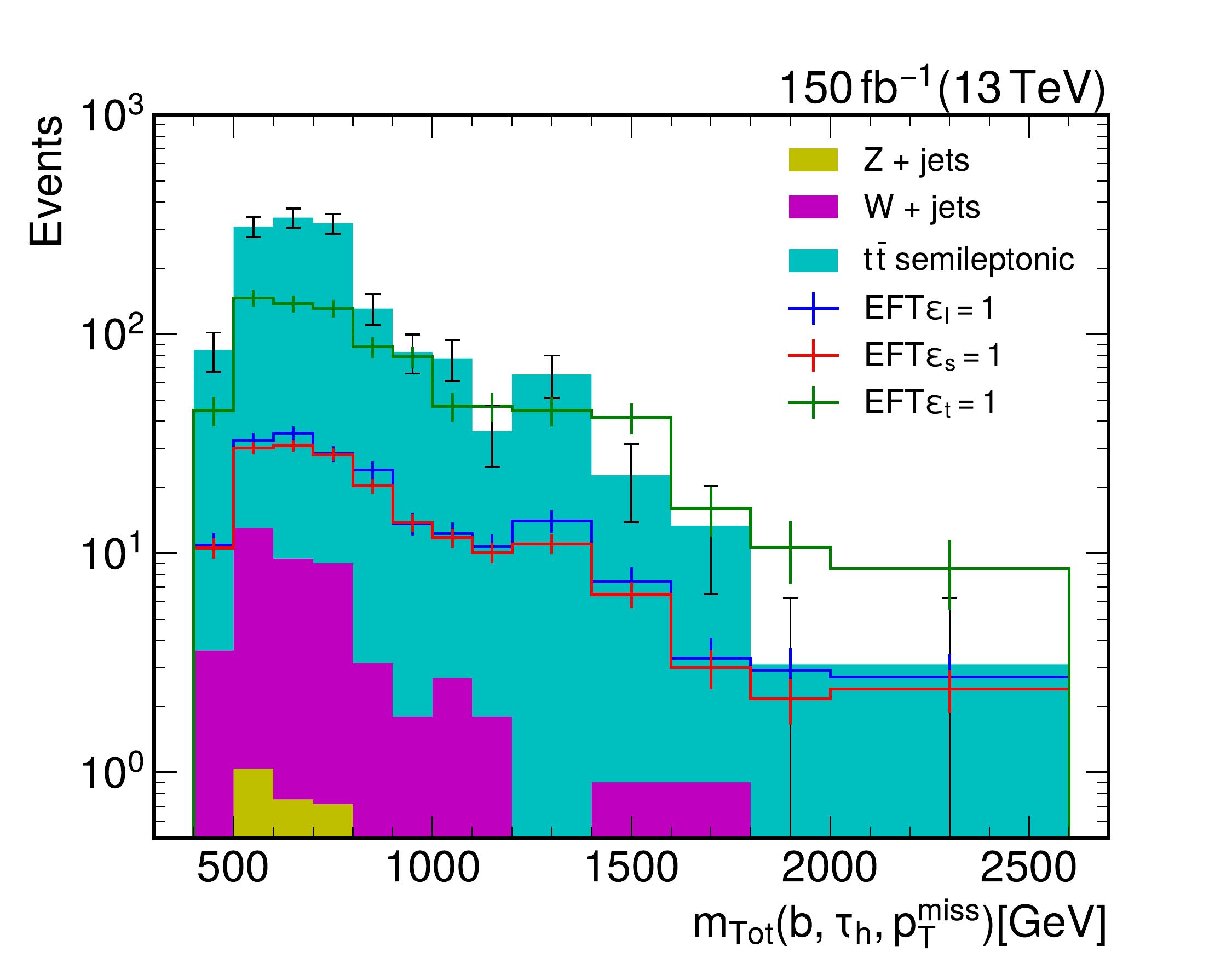}
\caption{Distribution of the total mass, $m_{\mathrm{Tot}}$, for the EFT benchmark scenario. The backgrounds are represented by solid colors and are stacked, while the signal samples are represented by solid lines and overlaid on top of the background. The expected event yields are estimated for an integrated luminosity of 150 $\mathrm{fb}^{-1}$ and $\sqrt{s} = 13$ $\mathrm{TeV}$.}
\label{fig:eftmass}
\end{figure}

Since the EFT is a point-like interaction, there is no on-shell mediator considered for these signals. In that sense, the interpretation of a peak in the histograms does not mean the same as a peak for the SSM samples. For this model, we use an observable, defined mathematically as $m_{\mathrm{Tot}}=\sqrt{\left(p_{\mathrm{T}}(\tau_h)+p_{\mathrm{T}}(b)+p_{\mathrm{T}}^{\text {miss }}\right)^2-\left(\mathbf{p}_{\mathrm{T}}(\tau_h)+\mathbf{p}_{\mathrm{T}}(b)+\mathbf{p}_{\mathrm{T}}^{\text {miss }}\right)^2}$ and named as the total mass, using the $\tau_{h}$, the $\mathrm{b}$-quark jet and the $p_{\mathrm{T}}^{miss}$. The distribution for background and three signal points is shown in Figure~\ref{fig:eftmass}.

Table \ref{tab:expnevents} shows the expected number of events, for three different masses, for the $\mathrm{W}'$ and $\mathrm{LQ}$ models and for three different couplings for the EFT interpretation, and for the dominant backgrounds.  

\begin{table*}[]
    \centering
    \begin{tabular}{ccc|cc}
    \cline{2-5}
    \multicolumn{1}{l}{}   & \multicolumn{2}{c}{$\sqrt{s} = 13\, \text{TeV}$} & \multicolumn{2}{|c}{$\sqrt{s} = 13.6\, \text{TeV}$} \\\cline{2-5}
                           & Baseline     & Topological   & Baseline       & Topological    \\ \hline\hline
    \multicolumn{5}{c}{SSM $\mathrm{W}'$}                                                                                       \\ \hline
    $m_{W'} = 0.6$ TeV     & $8.55\times10^4$    & $6.57\times10^3$    & $9.40\times 10^4$    & $7.22\times10^3$     \\
    $m_{W'} = 1.0$ TeV     & $7.78\times10^3$    & $2.46\times10^3$    & $8.55\times 10^3$    & $2.87\times10^3$     \\
    $m_{W'} = 1.6$ TeV     & $5.09\times10^2$    & $2.48\times10^2$    & $5.60\times10^2$     & $2.72\times10^2$     \\
    $m_{W'} = 2.0$ TeV     & $1.68\times 10^2$   & $8.73\times 10^1$   & $1.85\times 10^2$    & $9.63\times 10^1$    \\
    $m_{W'} = 2.5$ TeV     & $4.60\times 10^1$   & $2.30\times 10^1$   & $5.06\times 10^1$    & $2.53\times 10^1$    \\
    $m_{W'} = 3.0$ TeV     & $1.66\times 10^1$   & $7.25\times 10^0$   & $1.82\times 10^1$    & $7.97\times 10^0$    \\
    $t\bar{t}$             & $1.80\times 10^6$   & $1.11\times 10^3$   & $1.98\times 10^6$    & $1.22\times 10^3$    \\
    $W+Jets$               & $3.48\times 10^3$   & $2.86\times 10^1$   & $3.83\times 10^3$    & $3.14\times 10^1$    \\
    $Z+Jets$               & $7.41\times 10^1$   & $2.79\times 10^0$   & $8.70\times 10^1$    & $3.06\times 10^0$    \\ \hline
     \multicolumn{5}{c}{$\mathrm{LQ}$}                                                                                           \\ \hline
    $m_{LQ} = 0.6$ TeV     & $2.19\times10^4$    & $1.52\times10^2$    & $2.75\times10^4$     & $1.92\times10^2$     \\
    $m_{LQ} = 1.0$ TeV     & $5.56\times10^2$    & $1.24\times10^2$    & $7.01\times10^2$     & $1.55\times10^2$     \\
    $m_{LQ} = 1.25$ TeV    & $1.77\times10^2$    & $4.98\times10^1$    & $2.23\times10^2$     & $6.22\times10^1$     \\
    $m_{LQ} = 1.5$ TeV     & $5.42\times10^1$    & $1.79\times10^1$    & $6.84\times10^1$     & $2.25\times10^1$     \\
    $m_{LQ} = 2.0$ TeV     & $6.71\times10^0$    & $2.48\times10^0$    & $8.47\times10^0$     & $3.13\times10^0$     \\
    $m_{LQ} = 3.0$ TeV     & $1.82\times10^{-1}$ & $6.86\times10^{-2}$ & $2.29\times10^{-1}$  & $8.64\times10^{-2}$  \\
    $t\bar{t}$             & $1.80\times 10^6$   & $3.11\times10^2$    & $1.98\times 10^6$    & $3.42\times10^2$     \\
    $W+Jets$               & $3.48\times 10^3$   & $8.80\times10^0$    & $3.83\times 10^3$    & $9.82\times10^0$     \\
    $Z+Jets$               & $7.41\times 10^1$   & $7.16\times10^{-1}$ & $8.70\times 10^1$    & $7.87\times10^{-1}$  \\ \hline
    \multicolumn{5}{c}{EFT}                                                                                          \\ \hline
    EFT $\epsilon_{l} = 1$ & $1.52\times10^3$    & $1.94\times10^2$    & $1.71\times10^3$     & $2.18\times10^2$     \\
    EFT $\epsilon_{s} = 1$ & $9.09\times10^2$    & $1.81\times10^2$    & $9.99\times10^2$     & $1.99\times10^2$     \\
    EFT $\epsilon_{t} = 1$ & $1.09\times10^4$    & $8.41\times10^2$    & $1.31\times10^4$     & $1.01\times10^3$     \\
    $t\bar{t}$             & $1.80\times 10^6$   & $1.45\times 10^3$   & $1.98\times 10^6$    & $1.59\times 10^3$    \\
    $W+Jets$               & $3.48\times 10^3$   & $4.26\times 10^1$   & $3.83\times 10^3$    & $4.68\times 10^1$    \\
    $Z+Jets$               & $7.41\times 10^1$   & $3.69\times 10^0$   & $8.70\times 10^1$    & $4.05\times 10^0$    \\ \hline
   
    \end{tabular}
\caption{Expected number of events after the baseline and topological selection criteria, for the different signal models considered. The estimations are performed for $\sqrt{s} = 13$ and 13.6 $\mathrm{TeV}$, and 150 $\mathrm{fb}^{-1}$ luminosity.}
\label{tab:expnevents}
\end{table*}

To determine the discovery reach for each model and assess differences, we use a profile likelihood test statistic using the expected number of background and signal events in each bin of the distributions shown in Figures~\ref{fig:wprimemt},~\ref{fig:lqmass}, and~\ref{fig:eftmass}. We perform a maximum likelihood fit using the full range of these observables, employing a software package developed by the CERN laboratory known as ROOT-Fit \cite{Moneta:2010pm}. Systematic uncertainties are incorporated into the calculations as nuisance parameters, considering log-priors for normalization and Gaussian priors for shape uncertainties. The significance is calculated using the local $p$-value, estimated as the probability under a null signal hypothesis to obtain a value of the test statistic as large as that obtained with a signal plus background hypothesis. Similar to Refs.~\cite{Florez:2021zoo, Florez:2019tqr, Florez:2018ojp, Florez:2017xhf, VBFZprimePaper, Florez:2016lwi}, the  signal significance $\sigma_{\textrm{sig}}$ is obtained by calculating the value at which the integral of a Gaussian between $\sigma_{\textrm{sig}}$ and $\infty$ matches the local $p$-value. We consider a 3\% systematic uncertainty associated with the measurement of the integrated luminosity at the LHC, 5\% due to the PDF set used for the production of the simulated events for signal and MC, following the PDF4LHC prescription~\cite{Butterworth:2015oua}, 5\% on the identification of $\tau_{h}$ candidates and 5\% on the associated measurement of their energy scale, 3\% on $\mathrm{b}$-quark jet \cite{CMSbtag} identification efficiencies and a flat 10\% uncertainty to account for other sources of experimental effects, such as the resolution on the $p_{\mathrm{T}}^{miss}$ estimation.

\begin{figure}[]
    \centering
    \includegraphics[width=0.48\textwidth]{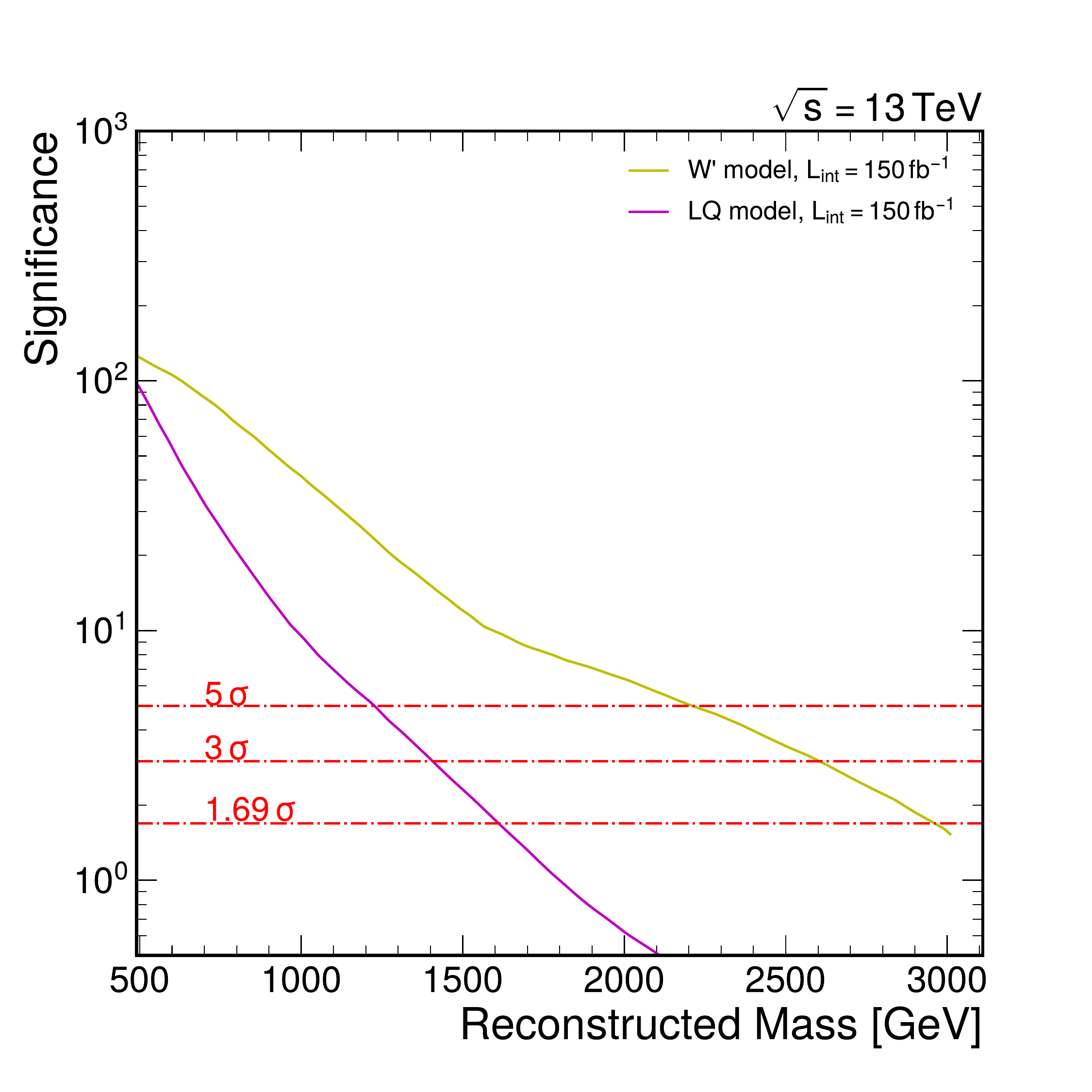}
    \caption{Expected signal significance for the $\mathrm{W}'$ and $\mathrm{LQ}$ benchmark scenarios, as a function of the corresponding reconstructed mass, considering an integrated luminosity of $150$ $\mathrm{fb}^{-1}$ and $\sqrt{s} = 13$ $\mathrm{TeV}$.}
    \label{fig:siglqwp}
\end{figure}

\begin{figure}[]
    \centering
    \includegraphics[width=0.48\textwidth]{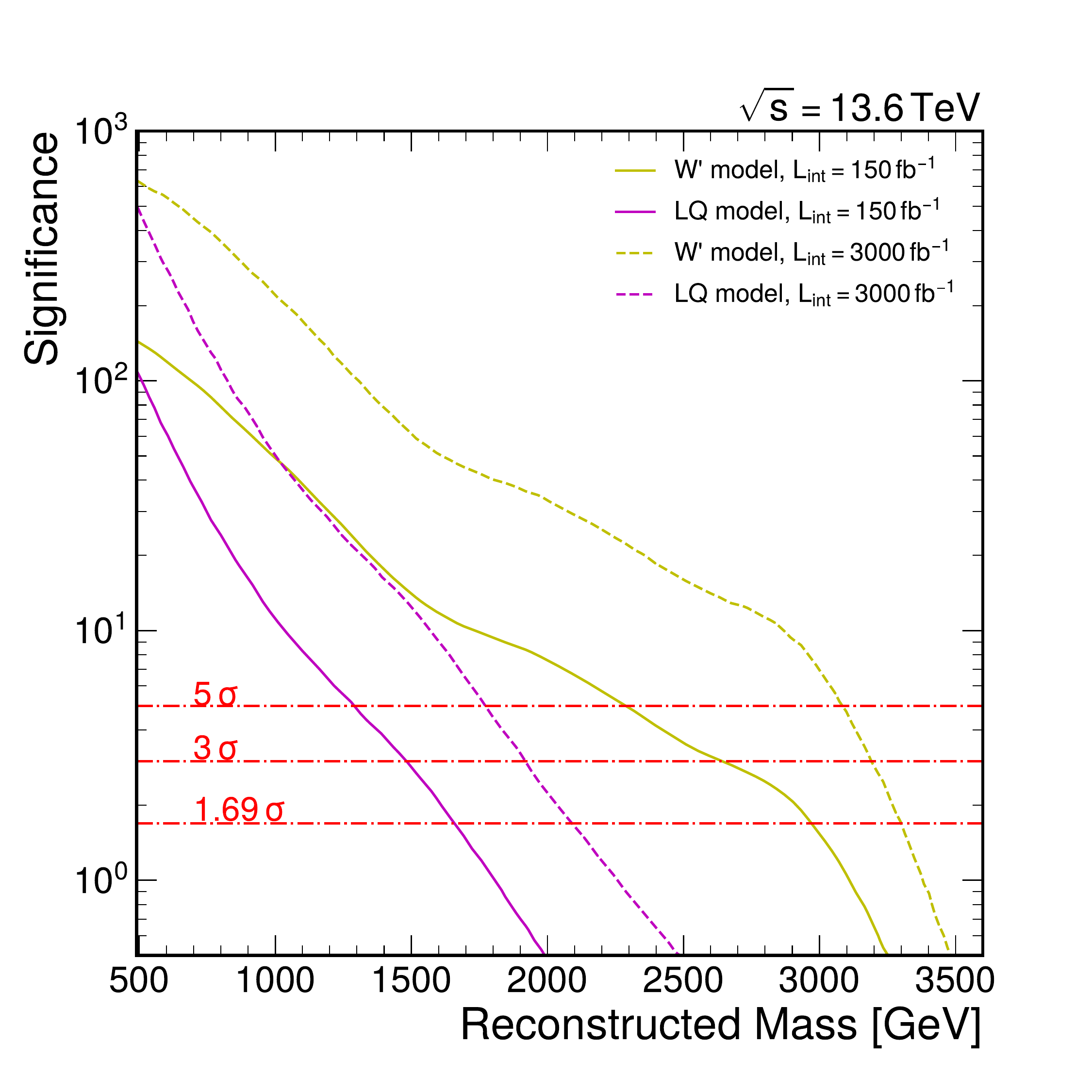}
    \caption{Expected signal significance for the $\mathrm{W}'$ and $\mathrm{LQ}$ benchmark scenarios, as function of the corresponding reconstructed mass, considering an integrated luminosity of $150 \mathrm{fb}^{-1}$ for $\sqrt{s} = 13$ TeV and $\sqrt{s} = 13.6$ $\mathrm{TeV}$, and $3000$ $\mathrm{fb}^{-1}$ for $\sqrt{s} = 13.6$ TeV.}
    \label{fig:siglqwp13p6TeV}
\end{figure}

Figure~\ref{fig:siglqwp} shows the expected signal significance for the $\mathrm{W}'$ ($\mathrm{LQ}$) model as a function of mass, for an integrated luminosity of 150 $\mathrm{fb}^{-1}$. For the SSM $\mathrm{W}'$ ($\mathrm{LQ}$) model, masses up to 2900 $\mathrm{GeV}$ (1600 $\mathrm{GeV}$) can be excluded considering 1.69$\sigma$ of signal significance, while masses up to 2300 $\mathrm{GeV}$ (1200 $\mathrm{GeV}$) could potentially be discovered with 5$\sigma$ signal significance above the background expectation. In addition, for the high luminosity LHC (HLL) at $\sqrt{s} = 13.6$ $\mathrm{TeV}$ and 3000 $\mathrm{fb}^{-1}$ luminosity, the projections show the extension of the exclusion reach to masses beyond 3 $\mathrm{TeV}$ for the SSM $\mathrm{W}'$ scenario, as shown in Figure~\ref{fig:siglqwp13p6TeV}. In the case of single $\mathrm{LQ}$ production, it is feasible to probe masses up to 2100 $\mathrm{GeV}$, while the sensitivity for discovery at 5$\sigma$ goes up to 1800 $\mathrm{GeV}$.  Table~\ref{tb:eftsign} shows the corresponding results for the EFT interpretation, for the three different coupling scenarios and LHC operation and luminosity conditions that have been considered.

\begin{table}[]
    \centering

\begin{tabular}{cccc}

    \hline
    EFT & \makecell{ $ \sqrt{s} = 13$ $\mathrm{TeV}$ \\ $(L_{int} = 150~ \mathrm{fb}^{-1})$ } & \makecell{ $ \sqrt{s} = 13.6$ $\mathrm{TeV}$ \\ $(L_{int} = 150~ \mathrm{fb}^{-1} )$} & \makecell{ $ \sqrt{s} = 13.6$  $\mathrm{TeV}$ \\ $(L_{int} = 3000~ \mathrm{fb}^{-1})$} \\
   \hline \hline
    $\epsilon_{S}$ & 2.3 & 2.4 &  10.6\\
    $\epsilon_{L}$ & 2.7 & 2.8 &  12.4\\
    $\epsilon_{T}$ & 9.2 & 10.2 & 45.4\\
    \hline
    \end{tabular}
    \label{tb:eftsign}
    \caption{Signal significance for the EFT benchmark scenario, considering integrated luminosities of $150 \mathrm{fb}^{-1}$, for $\sqrt{s} = 13$ TeV and $\sqrt{s} = 13.6$ TeV, and $3000 \mathrm{fb}^{-1}$ for $\sqrt{s} = 13.6$ TeV.}
\end{table}

\section{Discussion and conclusions}

Our findings have been interpreted in various models and therefore we have to discuss the implications in each of the scenarios. Firstly, while the current searches for a $\mathrm{W'}$ in the LHC, for the $\tau+p_{T}^{miss}$ final state quote exclusion limits up to 4.6~TeV~\cite{CMS-PAS-EXO-21-009,ATLAS:2018ihk} in the mass of the $\mathrm{W'}$, our proposal covers a new signature which must exist if a $\mathrm{W'}$ is the responsible for the anomalies on the $\mathrm{B}$ meson decay ratios. In this sense, our proposal pinpoints the specific physics process which would be a consequence of the observed $\mathrm{B}$ meson anomalies, while the generic searches for a $\mathrm{W'}$ cover a broader spectrum of processes. It is important to mention that the inclusion of a $\mathrm{b}$-quark jet in the final state changes significantly the expected amount and composition of the background, with respect to the inclusive search.  

In addition, the signature considered in this work is a novel signature to search for $\mathrm{LQ}$s. The closest final state considered in the literature for $\mathrm{LQ}$ searches, study final states with two $\tau_{h}$ and one $\mathrm{b}$-jet candidates. The exclusion limit achieved by that search sets a limit on the $\mathrm{LQ}$ mass, which should be greater than 1.1~$\mathrm{TeV}$~\cite{CMS-PAS-EXO-19-016}. Therefore, our proposal reaches higher and more stringent limit for the $\mathrm{LQ}$ mass and also explores a new channel relevant to study $\mathrm{B}$ meson anomalies. The gain in experimental sensitivity is associated with more optimal selection criteria, specially when considering only one $\tau_{h}$ candidate instead of two. The identification efficiency for $\tau_{h}$ candidates is not very high, on average 60\%.

\begin{figure}[]
    \centering
    \includegraphics[width=0.48\textwidth]{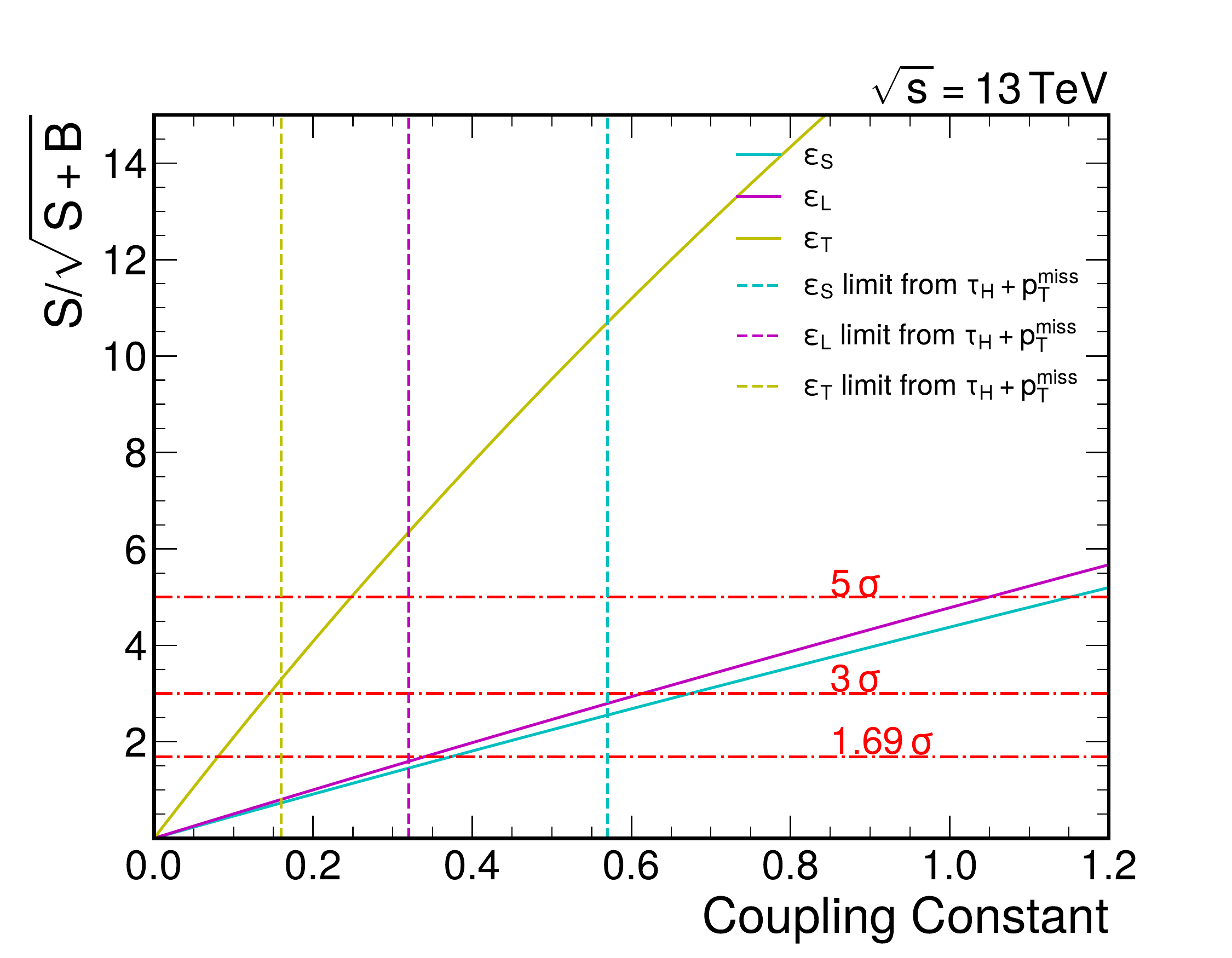}
    \caption{Expected signal significance for the EFT benchmark scenarios, as function of the corresponding strength of the interaction, considering an integrated luminosity of $150\, \mathrm{fb}^{-1}$ for $\sqrt{s} = 13$ TeV.}
    \label{fig:eftdiscu}
\end{figure}

For the EFT interpretation, we also reach new and complimentary results with regard to the state of the art. In Figure~\ref{fig:eftdiscu}, is shown the comparison among the limits achieved by our proposal and the limits derived in~\cite{greljo2019mono}. We can see that the limits using $b+\tau_{h}+p_{T}^{miss}$ signature are competitive with the $\tau_{h}+p_{T}^{miss}$ results. Additionally, from the physics point of view, both final states should exist if the assumed couplings exist and, therefore, our proposal makes a direct test of the underlying physics. In Table~\ref{tb:eftmorelim} are displayed the sensitivities, defined as $N_{S}/\sqrt{N_{S}+N_{B}}$,  that can be achieved for each EFT interpretation at the current operation energies for the LHC.

\begin{table}[]
    \centering

\begin{tabular}{rcc|cc|cc}
\cline{2-7}
\multicolumn{1}{l}{}                             & \multicolumn{2}{c|}{Z=1.69}      & \multicolumn{2}{c|}{Z=3}         & \multicolumn{2}{c}{Z=5}          \\ \hline
\multicolumn{1}{l|}{$\sqrt{s}$ $[\mathrm{TeV}]$} & \multicolumn{1}{c|}{13}   & 13.6 & \multicolumn{1}{c|}{13}   & 13.6 & \multicolumn{1}{c|}{13}   & 13.6 \\ \hline \hline
\multicolumn{1}{r|}{$\epsilon_{S}$}              & \multicolumn{1}{c|}{0.37} & 0.34 & \multicolumn{1}{c|}{0.67} & 0.61 & \multicolumn{1}{c|}{1.15} & 1.05 \\ \cline{2-7} 
\multicolumn{1}{r|}{$\epsilon_{L}$}              & \multicolumn{1}{c|}{0.34} & 0.31 & \multicolumn{1}{c|}{0.61} & 0.56 & \multicolumn{1}{c|}{1.05} & 0.95 \\ \cline{2-7} 
\multicolumn{1}{r|}{$\epsilon_{T}$}              & \multicolumn{1}{c|}{0.08} & 0.07 & \multicolumn{1}{c|}{0.14} & 0.12 & \multicolumn{1}{c|}{0.25} & 0.21 \\ \hline
\end{tabular}
\label{tb:eftmorelim}
\caption{Exclusion limits that can be achieved from this work for the EFT interpretation for various energies.}
\end{table}

Finally, we have developed a completely new strategy to address three different scenarios in a common final state taking into account the particularities of each model. In this sense, we have proposed a strategy that would be able to include information on the characteristics of an hypothetical signal in the LHC depending on the observable. In other words, we have shown how a signal in the considered final state does not have the same features for the possible theoretical models giving raise to it.

\begin{acknowledgments}

We wish to acknowledge the support of Ocampo-Henao who helped us with technical analysis tools. J.D.R.A. and T.A.G. gratefully acknowledge the support of the Colombian Science Ministry \-Min\-Cien\-cias and  Sos\-te\-ni\-bi\-li\-dad-UdeA. A.F thank the constant and enduring financial support received for this project from the faculty of science at Universidad de Los Andes (Bogot\'a, Colombia) and the Colombian Science Ministry \-Min\-Cien\-cias. 

\end{acknowledgments}

\appendix

\newpage
\bibliography{Paper}

\end{document}